# GAMIFICATION OF ENTERPRISE SYSTEMS – A SYNTHESIS OF MECHANICS, DYNAMICS, AND RISKS

*Manuel Schmidt-Kraepelin     Sebastian Lins     Scott Thiebes     Ali Sunyaev*
Research Group Critical Information Infrastructures,
Karlsruhe Institute of Technology, Germany
Website: http://cii.aifb.kit.edu

Abstract – Organizations highly depend on enterprise systems (ES), which are unlikely to develop their full potential if end-users neglect system usage. Accordingly, organizations attempt to overcome barriers to end-user acceptance in the ES context, which can be attributed to several factors on ES, organizational, and end-user level. Trying to take advantage of the growing passion for games, Gamification is a phenomenon proposed to motivate people by applying elements common to games in other contexts that have the potential to increase end-user acceptance. While first applications of Gamification exist in areas such as finance, health, and education, utility of gamifying ES has not been explored in-depth. Aiming to understand how Gamification can be applied to ES to increase user motivation, we analyze literature concerning game elements (i.e., mechanics and dynamics) used in Gamification and related risks. Our study yields a synthesis of mechanics in clusters of system design, challenges, rewards, and user specifics as well as related dynamics. We discuss the extent to which the game elements can be used to address ES acceptance barriers. While our study reveals that Gamification has potential for motivating ES users, future research should analyze concrete implementations of Gamification in ES contexts to investigate long-term effects.

## 1. Introduction

Organizations heavily invest into enterprise systems (ES) (Majchrzak et al., 2009; Kim et al., 2010; Tong et al., 2015) for purposes such as boundary-less structures, streamlined business processes, integrated systems, and non-redundant managerial tasks (Davenport, 1998; Grant, 2003; McAfee, 2009). However, the majority of projects to implement an ES fail due to abandonment of the systems, unfulfilled benefits, or substantial financial drawbacks (Wu, 2011; Tong et al., 2015).

In their continuous attempt to explore how organizations can turn their ES investments into performance, researchers have paid particular attention to end-users (Ke *et al.*, 2012). Successful ES require end-users to accept organizational and technical changes and to use the ES (Wagner *et al.*, 2012). Understanding factors that drive end-user acceptance concerning (non-)technical changes is essential for effective ES usage (Wu, 2011; Chou *et al.*, 2014). Whereas promotion of ES is thus critical for end-user acceptance, research so far provides insufficient guidance for end-user engagement (Nwankpa and Roumani, 2014b). In other words, actions and interventions to best promote ES usage remain unclear (Nwankpa, 2015).

As an approach to overcome these challenges, Gamification (see "Gamification: Basic Concept" for a detailed definition and exemplary illustration) has been proposed to improve ES utilization (Schacht and Schacht, 2012; Raftopoulos, 2014). However, Gamification's potential for and threats to ES have not been explored yet. In order to address this research gap, we answer the following research questions:

1. *How can Gamification be applied to increase ES end-user acceptance?*
2. *What risks are related to such applications of Gamification in ES contexts?*

Based on a comprehensive literature review, we identify and analyze game elements to be used to gamify ES as well as risks associated with such Gamification. Drawing on these findings, we discuss how Gamification can be applied to ES in order to overcome prevailing problems of end-user acceptance. With our study, we pave the way for applications of Gamification in the IS domain and provide a two-fold contribution. First, we advance the understanding of Gamification by providing a synthesis of previous research in this novel field. Second, we guide future research as well as business applications towards an effective usage of game elements in ES while considering related risks.

This article proceeds as follows. Next, we describe the evolution and challenges of ES, explain the basic concept of Gamification as well as its relevance for the IS research domain, present an example of a gamified ES function, and introduce different types of game elements. Subsequently, we outline the research approach and describe the process of identifying and analyzing relevant literature. Our article continues by presenting the results concerning game elements and describing risks associated with ES gamification. We finally discuss our findings in light of previous ES acceptance research, give practical implications of our findings, and discuss limitations and directions for future research. Our study ends with a conclusion.

## 2. Theoretical Background

### 2.1 Enterprise System Acceptance

ES are organization-spanning standardized software systems that integrate applications and data into a single instance to optimize information flows in organizations (Davenport, 1998; Moon, 2007). For ES implementations to be successful, organizations need to ensure that end-users learn to accept organizational and technical changes and to use the system (Wagner *et al.*, 2012). In general, lack of user acceptance is seen as a major reason for failure of innovative information technologies (Venkatesh *et al.*, 2003). In particular, inadequate understanding of ES is a major reason for their limited acceptance and end-users creating and using workarounds that limit systems' effectiveness (Nwankpa and Roumani, 2014a). Understanding what factors drive end-users to accept the (non-)technical changes and to grapple with necessary knowledge and skills is thus essential for effective ES usage (Wu, 2011; Chou *et al.*, 2014). Explanations for how and why end-users accept ES are diverse can be attributed to three levels: ES characteristics, characteristics of the organization, and characteristics of the end-users.

### 2.2 Enterprise System Characteristics

The fit between ES and organizational requirements is considered important for ES acceptance (See-Pui Ng, 2013). High degrees of organizational fit, which comprises data fit, process fit, and user interface fit, positively impact end-user satisfaction and thus acceptance (See-Pui Ng, 2013; Nwankpa, 2015). Additionally, the complexity of ES is found to impact usage behavior. End-users causally attribute system complexity to continued negative performance outcomes, which leads to decreased self-efficacy towards the ES (Kelley *et al.*, 2013). Finally, the very nature of ES can impact usage behavior since they automate business processes and prescribe user actions. As a consequence, end-users need to alter the way they previously performed tasks and to comply with inflexible business rules. Hence, it is less likely that they develop an enhanced usage behavior (Bagayogo *et al.*, 2014).

Organizations need to help end-users understand ES by providing training and enabling post-implementation learning (Doll *et al.*, 2003; Chang and Chou, 2011; Chou *et al.*, 2014). The complexity of an ES limits the amount of knowledge users can acquire through training prior to the implementation, thus resulting in a gap between how a system is actually used and the understanding of its full potential (Cooper and Zmud, 1990; Yi and Davis, 2003; Chang and Chou, 2011). While training is often provided prior to the implementation of a system and aims at enhancing initial usage, post-implementation learning lets end-users practice with the actual system after the implementation (Doll *et al.*, 2003; Chang and Chou, 2011). Post-implementation learning helps bridging this gap and plays a central role for realizing the full potential of ES (Cooper and Zmud, 1990). Furthermore, an important antecedent of ES acceptance can be traced to the working environment. For example, high task interdependency can lead to adaption of processes and user behavior (Bagayogo *et al.*, 2014). In line with this finding, a high level of task variety helps to motivate users in order to find new and creative ways to solve prevalent problems in regard to the ES usage (Liang *et al.*, 2015). However, coming up with new and creative ways of solving problems requires high job autonomy. Thereby, users can self-explore features and develop novel forms of use (Ke *et al.*, 2012; Liang *et al.*, 2015). Especially during the early post-implementation phase, users experience a steep learning curve and might even struggle with simple tasks. When they seek to understand the new system, organizational support is needed to facilitate desired behavior (Boudreau and Seligman, 2005; Tong *et al.*, 2015). If this sense making process of users is not managed properly, user resistance might eventually lead to system abandonment. Organizational support also relies on the social network within an organization (e.g. contacting IS specialists or asking peer users for help) to provide guidance during this phase of insecurity and impersonal support such as documentations (Tong *et al.*, 2015). Organizational culture can create a social system that lets individuals recognize the value of their tasks and duties by exchanging and discussing issues related to the ES with fellow end-users (Ke *et al.*, 2012). Organizational culture also allows end-users to apply novel ways to use ES (Liang *et al.*, 2015).

Several researchers relate characteristics of individuals to the process of ES acceptance. Culture is seen as an important moderator in this context (Alhirz and Sajeev, 2015). For instance, the importance of factors for ES acceptance differs between eastern and western cultures (Hwang, 2012). Specifically, innovativeness as predictor for the intention to use is important in cultural environments that are characterized by high levels of uncertainty (i.e., eastern cultures), while individuals' intrinsic motivation as an antecedent of intention to use is more important in western cultures. Besides cultural factors, the extent to which users are intrinsically motivated is found to impact system use. Intrinsic motivation to use a system is a requirement for organizations to realize benefits from ES implementations since it leads to the discovery of innovative ways of usage that best support user tasks (Seddon *et al.*, 2010; Ke *et al.*, 2012). This is especially important when considering ES complexity, which requires significant cognitive efforts to achieve adaptive system use (Sharma and Yetton, 2003; Sykes *et al.*, 2009; Ke *et al.*, 2012). Intrinsically motivated users put a lot of effort in gathering information and developing novel ways of ES usage, which leads to exploratory usage (Ke *et al.*, 2012). Furthermore, since intrinsically motivated individuals enjoy their work as such, they are also likely to derive enjoyment from exploring ES, which leads to exploration satisfaction (Ke *et al.*, 2012; Nwankpa and Roumani, 2014a).

Researchers have paid particular attention to end-users in their continuous attempt to explore how organizations can turn their ES investments into performance (Ke *et al.*, 2012). In this context, recent research emphasizes the criticality of the early post-adoptive phase, that is, the phase subsequent to deploying the system in the organization (Schwarz *et al.*, 2014; Tong *et al.*, 2015). Rather than frequency or duration of system usage, end-users' behavior concerning the incorporation of the ES into work routines is decisive for organizational performance (Schwarz *et al.*, 2014) and intrinsic rather than extrinsic motivation is key to ES acceptance (Ke *et al.*, 2012). Accordingly, promotion of ES usage is crucial for organizations. However, previous research provides insufficient guidance for managers in organizations who must grapple with this dilemma (Nwankpa and Roumani, 2014b) and the managerial actions and interventions that can best promote ES usage remain unclear (Nwankpa, 2015). As an approach to change behavior through addressing intrinsic motivation, Gamification, which offer benefits of both utilitarian and hedonic information systems (Hamari and Koivisto, 2015), is an approach to be applied to ES in order to overcome these challenges (Schacht and Schacht, 2012; Raftopoulos, 2014).

## 2.3 Gamification: Basic Concept

Trying to take advantage of the growing passion for games (McGonigal, 2011), Gamification is a trending phenomenon that aims to motivate people by applying elements common to games in different contexts (Deterding *et al.*, 2011a). Originating from the digital media domain (Deterding *et al.*, 2011b), Gamification experienced a widespread adoption, which is evidenced by the emerging trend on scientific publications and general search queries in the research field of Gamification (see Figure 1). While we acknowledge that other definitions exist, we align our understanding of Gamification[1] with the definition by Deterding *et al.* (2011a): "the use of game design elements in non-game contexts". By following this broad definition, we are able to include the majority of research for our analysis.

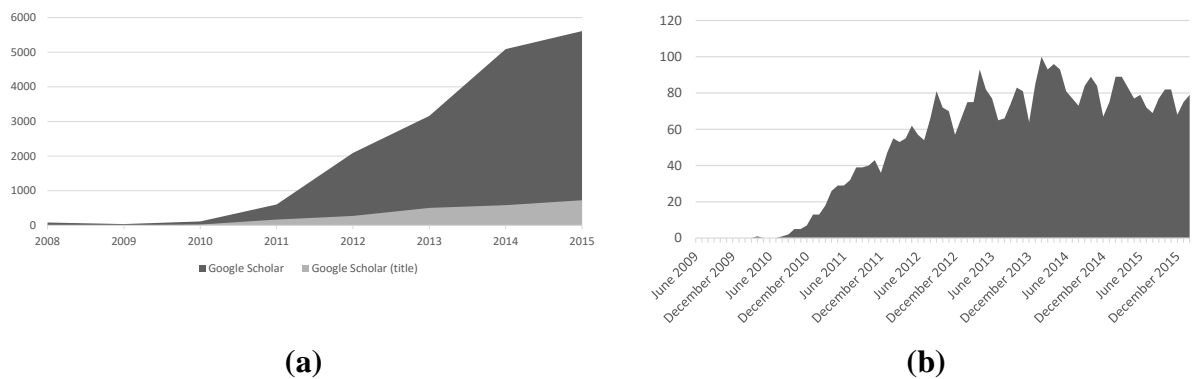

(a)          (b)

**Figure 1** Search hits for "gamification" (a) and Google Trends for "gamification" (monthly) (b).

To exemplify Gamification's applicability in the ES context, we describe the use of selected game elements to foster user engagement concerning a customer relationship management (CRM) module, which is considered a common feature of ES (Chen and

---

[1] Akin paradigms include pervasive and serious games. While pervasive games expand the borders of traditional digital games in a spatial, temporal, or social manner (Montola *et al.*, 2009), serious games are full-fledged games for non-entertainment purposes. Instead, Gamification uses game elements only (Deterding *et al.*, 2011a).

Popovich, 2003; Chalmeta, 2006; Tarantilis *et al.*, 2008). Figure 2 shows a possible dashboard for CRM employees (cf. the open source solution Zurmo for gamifying CRM; http://zurmo.org/), which is based on common game elements[2]. One of the CRM module's aims is to increase *competition* among employees, for instance, by using the following game elements (henceforth we refer to this case as the 'CRM example'). Employees receive *badges* (left side of Figure 2) for accomplishing specific *goals*. For instance, employees are fostered to use the system by receiving badges when a specified number of logins is reached (e.g., 25 logins). Other *badges* can be earned by fulfilling *goals* like conducting contact searches or creating new products. For accomplishing tasks in diverging areas of the CRM system, users receive *points* (lower right of Figure 2). Based on the amount of points gained, users reach specific *user levels* (lower left of Figure 2; e.g., level 1 requires at least 200 points, level 2 500 points, level 3 1,000 points and so forth). Furthermore, the design of the CRM system shown in Figure 2 includes a *leaderboard* (upper right of Figure 2), which shows a user's rank according to the amount of awarded points in comparison to other users. Referring to aforementioned game elements, Figure 3 shows a possible interaction between a CRM user and the gamified CRM system.

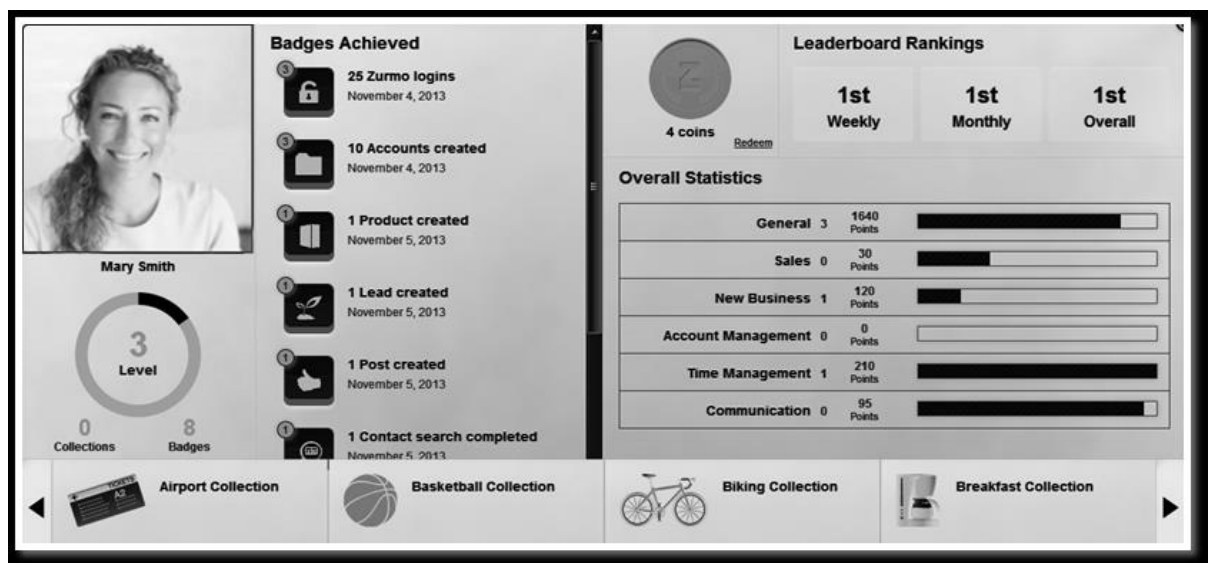

**Figure 2** Example of a gamified CRM (retrieved from http://zurmo.org/features/gamification-phase-ii-is-here; image colors have been inverted for improved readability).

---

[2] The game elements used in this example are highlighted in italics. While definitions can be found in Appendix A, we describe these game elements in more detail within our results section.

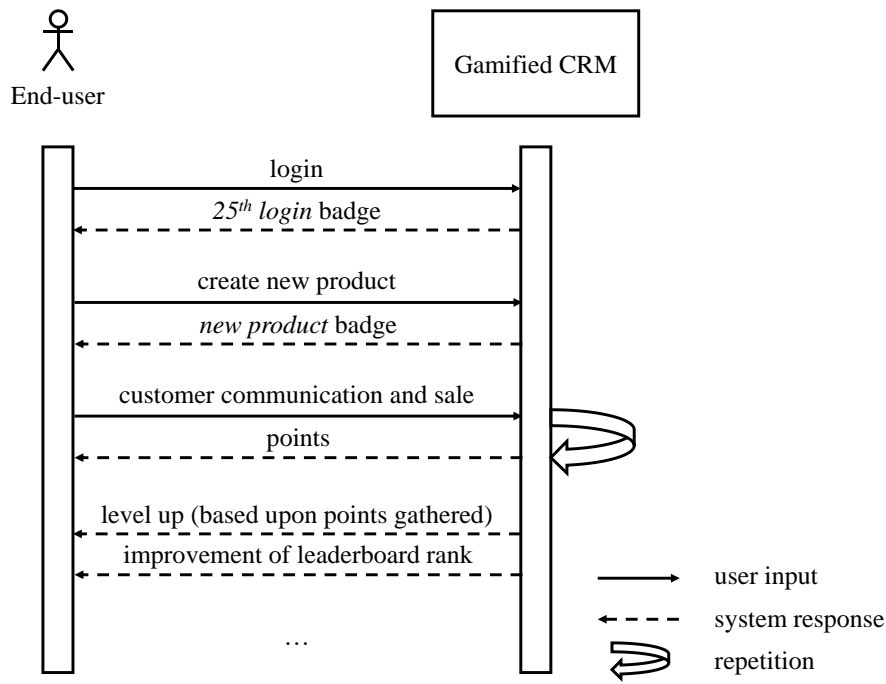

**Figure 3** Illustration of exemplary interaction between CRM user and gamified CRM system.

In the example described above, the game elements are used to evoke employees to use the CRM module in order to improve the organization's effectiveness. In its most common sense, Gamification uses game elements to regulate people's behavior in non-game activities for strategic purposes (Schrape, 2014). In contrast to modern marketing techniques that aim at changing the way people think (e.g., people's attitudes and beliefs), Gamification aims at changing the way people behave (Schrape, 2014). For this purpose, Gamification utilizes two interrelated aspects of today's world. First, video games have become a substantial part of daily life (McGonigal, 2011). Second, video games "can demonstrably produce states of desirable experience, and motivate users to remain engaged in an activity with unparalleled intensity and duration" due to video games' explicit design focus on entertainment rather than utility (Deterding *et al.*, 2011b: 2). Accordingly, Gamification can be used to make non-game applications (e.g., the CRM module) more motivating and engaging to use. As empirical findings indicate, the incorporation of game elements into repetitive and monotone tasks makes them more fun and enjoyable (Flatla *et al.*, 2011). Gamification also increases user participation and deepens user involvement (Rapp *et al.*, 2012; Barata *et al.*, 2013).

Gamification's positive influence on motivation and user behavior is what makes this concept highly relevant for IS research. Different approaches trying to explain system acceptance and usage exist in research, many of which having in common the individuals' opinions and perceptions of IS as an important factor for their behavior (Agarwal and Karahanna, 2000). Moreover, it is typically argued that so-called intrinsic factors are essential for motivating people effectively and sustainably (Ke *et al.*, 2012).

While little research exists concerning how Gamification can be successfully used within intra-organizational contexts (Fitz-Walter *et al.*, 2012; Huotari and Hamari, 2012; Hamari, 2013), trying to foster effective usage of ES through the application of Gamification bears high potential to counteract the growing complexity and scope of such systems (Schacht and Schacht, 2012; Sommerville *et al.*, 2012; Raftopoulos, 2014). It can be applied in context of ES to offer a new research stream concerning ES user acceptance (Hamari, 2013). Considering that future employees will have grown up in a world where video games are common and available to everyone, using Gamification as means to motivate them seems reasonable (Burke and Hiltbrand, 2011).

Positive aspects notwithstanding, concerns related to the effects of Gamification are increasingly being raised. Critics say that Gamification is a buzzword used by companies as a mere marketing tool (Chorney, 2013) and that many people will not respond to Gamification (Spencer, 2013). They also argue that Gamification might induce unwanted behavior if game elements become more important than the actual core function (Hakulinen *et al.*, 2013; Haaranen *et al.*, 2014). It is thus crucial to identify the contexts in which Gamification will be particularly useful (Blohm and Leimeister, 2013). Hence, we dedicate this study to analyze how Gamification can be generally applied to ES.

## 2.4    Gamification: Mechanics & Dynamics

While the previous section has introduced the basic concept of Gamification and given an example of how to use game elements in an ES context, we now take a closer look at different types of game elements and their interrelatedness. In this context, research has identified two major types of game elements, which are commonly referred to as game mechanics and game dynamics (short: M&Ds) (Zichermann and Cunningham, 2011).

Mechanics, on the one hand, are functional components of a gamified application and provide various actions and control mechanisms to enable user interaction (Hunicke *et al.*, 2004). As illustrated in Figure 2, common mechanics include point systems, leaderboards, levels, and goals (Zichermann and Cunningham, 2011). Mechanics can thereby be categorized into the following four clusters (see Table 1 for an overview). Mechanics related to *system design* refer to the way gamified applications are developed and designed in order to increase motivation of end-users and thus to enhance their engagement. An example is the implementation of feedback in form of progress bars (Huotari and Hamari, 2012). *Challenges* are mechanisms used to inspire employees to accomplish specific tasks. Setting clear goals for specific tasks can be considered a typical example for this type of mechanisms (Passos *et al.*, 2011). Related to the accomplishment of tasks is the granting of *rewards* (Hamari, 2013), which, for instance, reflect users' successfulness in solving challenges. Moreover, *user specifics* are applied to enhance motivation through the influence on individual personality. Such influence can, for instance, be realized by providing users the opportunity to design and use virtual characters (Barata *et al.*, 2013).

**Table 1** Clusters of Mechanics.

| Cluster | Definition |
|---|---|
| System Design | Mechanics describing how a gamified application has to be designed and developed to motivate users. |
| Challenges | Mechanics attempting to motivate users by providing challenges as well as mechanics supporting the development or accomplishment of challenges. |
| Rewards | Mechanics aiming to motivate users by providing after certain actions were successfully taken. |
| User Specifics | Mechanics aiming to motivate users by directly influencing the individual personality. |

Dynamics, on the other hand, determine an individual's reactions as a response to implemented mechanics. These reactions try to satisfy fundamental needs and desires, including the desire for reward, self-expression, and altruism. Moreover and concerning the CRM example, competition is a dynamic that might result from users striving for high ranks in leaderboards. Thus, adequate combinations of M&Ds create a motivating, emotional, and entertaining interaction (Neeli, 2012).

While various M&Ds have been applied to previous empirical studies (e.g., Burke and Hiltbrand (2011), Liu *et al.* (2011), and (Vassileva, 2012)), a comprehensive overview of M&Ds is provided in white papers only and is currently lacking in the academic discourse. For future applications of Gamification, Hamari *et al.* (2014) call for more rigorous methodologies that pay increased attention to the context being gamified and the qualities of the users. More rigorous studies require researchers to choose Gamification techniques that fit the context (Hamari *et al.*, 2014). One premise for choosing an adequate technique is the awareness of diverse M&Ds to be used in order to increase user engagement.

Concluding, research lacks guidance for ES end-user acceptance, which is required to benefit from the substantial investments into such systems. Researchers have suggested to gamify ES as a potential solution for this challenge. However, applying Gamification to ES requires a comprehensive overview of the diverse options and related risks, which we provide in this paper based on a literature review.

## 3. Research Approach

We applied a two-step research approach (see Figure 4), first conducting a literature review to identify relevant publications, and then analyzing the identified publications in view of M&Ds and risks. The process of reviewing and coding literature was adapted from Lacity *et al.* (2010) and Jeyaraj *et al.* (2006).

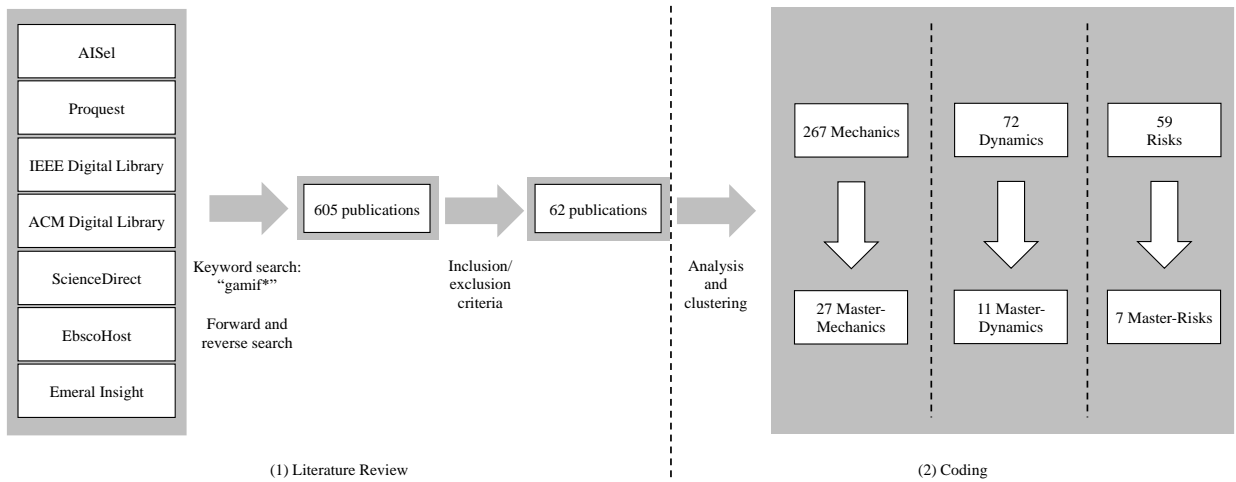

**Figure 4** Research approach for identification and analysis of Gamification M&Ds and risks.

## 3.1 Literature Review

Our descriptive literature review (Paré *et al.*, 2015) was guided by recommendations for reviews in the software and information systems domain (Webster and Watson, 2002; Kitchenham, 2007; vom Brocke *et al.*, 2015). In the following, we describe and argue for the single steps applied, while Figure 5 illustrates our literature selection process.

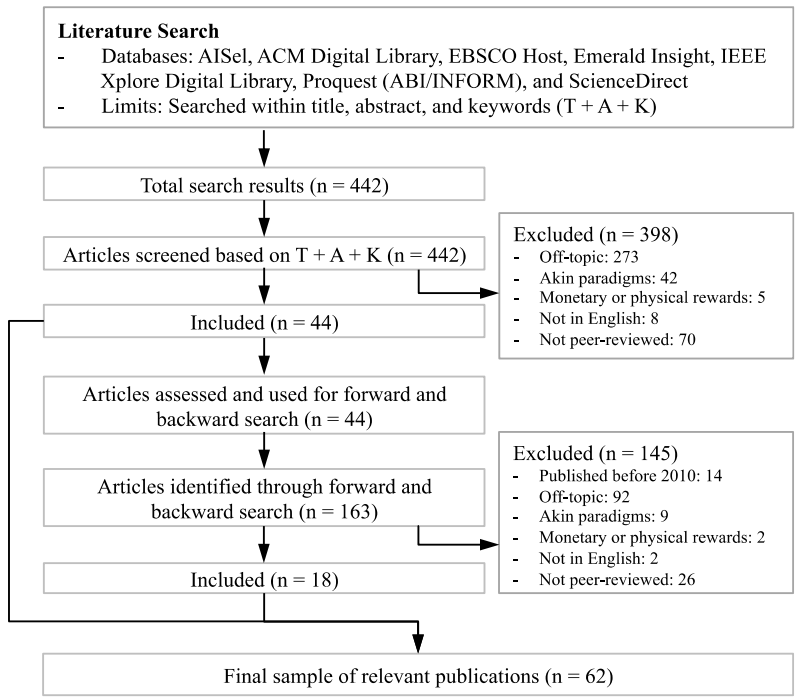

**Figure 5** Literature Selection Process.

For the identification of papers addressing Gamification, we searched scientific databases that we deemed representative as they cover a wide range of journal articles as well as conference publications: Association for Information Systems Electronic Library (AISel), Association for Computing Machinery (ACM) Digital Library, EBSCO Host, Emerald Insight, Institute of Electrical and Electronic Engineers (IEEE) Xplore Digital Library,

Proquest (ABI/INFORM), and ScienceDirect. To cover a broad set of publications, potentially relevant papers needed to meet the "gamif*" search string in title, keywords, or abstract. We limited our search to sources published after 2009 since the term 'Gamification' did not widely diffuse until the second half of 2010 (the search was performed in June 2015). Our search yielded a total of 442 potentially relevant articles, of which we considered 44 for detailed analysis. Additionally, we identified 163 potentially relevant articles through searching forward (i.e., using Google Scholar) backward (Webster and Watson, 2002).

Due to the novelty of Gamification, a restriction to publications in highly ranked journals, conferences, or magazines seemed inappropriate. Therefore, two authors separately reviewed identified studies in detail and assessed their relevance for this study. Predominantly, we included empirical studies on Gamification. To ensure applicability of identified M&Ds to the ES context, we also included conceptual studies addressing Gamification in workplace settings. Moreover, we excluded studies in which external rewards (e.g., monetary or physical) were awarded for high participation since Gamification should be applied to foster intrinsic motivation, deepen system involvement, and to create gameful and fun experiences on a long-term basis (Huotari and Hamari, 2012; Hamari, 2013). Empirical studies were included in order to gain insights into potential Gamification implementations and to gather information about empirical evaluations of applied M&Ds. After individual classifications were completed, two authors compared and discussed their results. Out of the 605 remaining studies, 62 were declared as relevant for this research according to the criteria listed in Table 2. These studies are marked with asterisks in the references. Detailed information concerning the exclusion of studies are provided in Figure 5.

**Table 2** Inclusion and exclusion criteria.

| Inclusion Criteria | Exclusion Criteria |
|---|---|
| Published after 2009 | Published before 2010 |
| Focus on Gamification | No direct connection to Gamification (e.g., akin paradigms; "Gamification: Basic Concept") |
| Empirical studies on Gamification **and/or** workplace as study setting | Conceptual studies that do not relate to a workplace setting (e.g., education, health, crowd sourcing) |
|  | Study contexts with monetary or physical rewards |
|  | Editorials, papers not written in English language |

### 3.2   Data Analysis

Our data analysis follows the approach by Jeyaraj *et al.* (2006). As a first step, we carefully read and analyzed all relevant studies to identify Gamification M&Ds. Furthermore, we derived a name and a description for all identified M&Ds based on the information provided in the respective articles. In order to uniformly code the influence of the M&Ds on the motivation in qualitative and quantitative studies, we adapted the coding scheme by Jeyaraj *et al.* (2006), which assigned three possible values to the influence: '+', '-', or 'o'. The following rules were adopted for coding the presumed influence of M&Ds on motivation (see also Table 3). A positive influence of a mechanic or dynamic on motivation was coded as a

'+', if a positive empirical confirmation was given (in case of quantitative studies) or the authors strongly argued (in case of qualitative studies) that a positive influence was exerted. Analogously, a negative influence of mechanics or dynamics on motivation was coded as a '-', if a negative empirical confirmation was given (in case of quantitative studies) or the authors strongly argued (in case of qualitative studies) that a negative influence was exerted. Otherwise, no influence of mechanics or dynamics on motivation was coded as a 'o', if no empirical confirmation was given in the source (in case of quantitative studies) or the authors strongly argued (in case of qualitative studies) against the existence of an influence on motivation.

**Table 3** Coding values adopted from Jeyaraj *et al.* (2006).

| Coding | Description |
|---|---|
| + | An empirical confirmation was given (quantitative studies) or the authors strongly argued (qualitative studies) that a positive influence was exerted. |
| - | An empirical confirmation was given (quantitative studies) or the authors strongly argued (qualitative studies) that a negative influence was exerted. |
| O | No empirical confirmation was given in the source (quantitative studies) or the authors strongly argued (qualitative studies) against the existence of an influence on motivation. |

To aggregate the identified M&Ds, we adopted the method of Lacity *et al.* (2010). Lists of so-called Master-Mechanics (Master-Ms) and Master-Dynamics (Master-Ds; we use 'Master-M&Ds' for cases in which we refer to both types of game elements) were created. A Master-M or Master-D is an aggregation of similar mechanics or dynamics, respectively, consisting of a name and a description (see the bootstrapping approach in Jankowicz (2004)). If an identified dynamic fitted into an existing Master-D, we assigned it accordingly; otherwise, a new Master-D was created. During this process, we avoided semantic ambiguities as suggested by Shaw and Gaines (1989). Since different people often put the same labels on different things and vice versa, it is crucial for the validity of a qualitative analysis to be aware of potential semantic ambiguities. Shaw and Gaines (1989) mention four possible semantic constellations: consensus (the same terminology is used for same concepts), correspondence (different terminology for same concepts), conflict (same terminology for different concepts), and contrast (different terminology for different concepts), for which we attributed in our analysis. Completing the analysis, we finalized the Master-Ds list by reviewing all assignments. We applied the same approach for mechanics in relation to Master-Ms and additionally classified the Master-Ms according to the clusters in Table 1.

We were able to identify 339 M&Ds in total, of which 172 were empirically confirmed to exert influence on motivation. Furthermore, a positive influence on motivation was coded 309 times, no influence was coded 23 times, and negative influence was coded seven times. These 339 M&Ds were then matched to 38 Master-M&Ds (see the 27 Master-Ms and 11 Master-Ds in Figure 6 along with the number of articles in which the Master-M&Ds have been coded; all Master-M&Ds with their descriptions are listed in Appendix A and Appendix B).

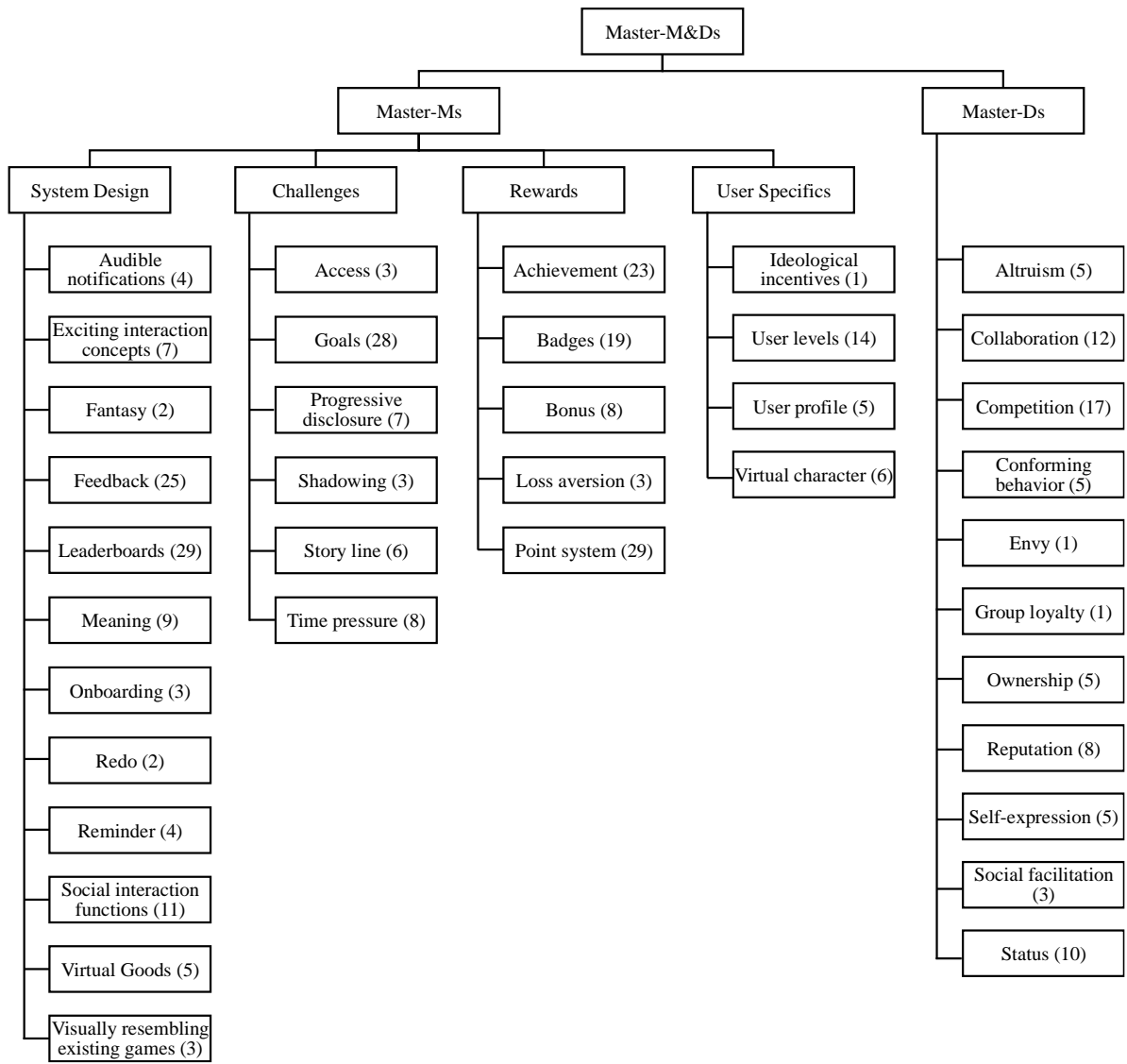

**Figure 6** Identified Master-M&Ds with their number of times being coded.

To ensure that we identified a reliable set of Master-M&Ds, we followed researchers stressing that an important goal is to reach theoretical saturation (Glaser and Strauss, 1967; Strauss and Corbin, 1990) regarding the emerging Master-M&Ds, that is, the point when no new findings are gained in further articles. Lincoln and Guba (1985: 235) speak of the term 'point of redundancy' in this context. Since no new Master-M&Ds emerged in the last 19 articles identified in our literature review, we are confident to have reached saturation.

To address the research gap concerning risks of Gamification, we further extracted perceived risks, that is, potentially negative consequences of gamified ES. This extraction led to 59 risks, which were aggregated in the same manner as M&Ds. This aggregation led to seven Master-Risks.

## 4. Results

Our results are divided into three subsections. First, we describe the Master-Ms of Gamification. Second, we explain the Master-Ds that might result from using ES gamified with the Master-Ms. Finally, we refer to the risks that need to be addressed when gamifying ES.

### 4.1 Gamification Mechanics

We describe the Master-Ms according to their mapping to one of four clusters: system design, challenges, rewards, and user specifics (see Table 1). For each Master-M, we provide an explanation as well as examples of its application and refer to combinations with related Master-M&Ds.

*System Design*

The two most often coded Master-Ms of this category are *leaderboards* and *feedback*. In the context of Gamification, leaderboards are used to track and display action progress. Leaderboards usually motivate users for two reasons. First, they make one's personal performance visible and present it to others (i.e., showing status), thus demonstrating one's capabilities. Second, they promote competition among participating users, which is sustained through regular leaderboard updates (Depura and Garg, 2012; Gordillo *et al.*, 2013). In the ES context, the application of leaderboards can be suitable since existing organizational structures can be used to foster competition (Yates and Wootton, 2012). Leaderboards can, for example, be developed for individuals, teams, organizational units, and different locations. In Figure 2, the leaderboards differentiate between different time periods (ranks for current week, current month, and overall). Accordingly, new users are also able to compete for the first rank. With respect to the design of gamified ES, another important Master-M is *feedback*. Feedback should be immediate and motivating, especially after an individual action was completed (Wang and Sun, 2011; Groh, 2012). Users should be informed about doing something wrong and be allowed to undo false actions (Sweetser and Wyeth, 2005). Moreover, users should be permitted to *redo* unsuccessful actions or to achieve sub-goals (Tootell *et al.*, 2014). A simple example of an ES feedback mechanism is the use of progress bars, which indicate the progress of filling in a form and inform users about false inputs (Huotari and Hamari, 2012). Progress bars can create a feeling of achievement, just by performing several actions (Burke and Hiltbrand, 2011). Additionally, non-disturbing and short-timed messages (Meder *et al.*, 2014) or windows can be shown, for example, in one corner of the screen (Denny, 2013; Meder *et al.*, 2013). Furthermore, knowledge maps are common means to show an overall map or tree structure to guide users in accomplishing the next task (Morrison and DiSalvo, 2014). Each feedback mechanism has to be evaluated in the working context, such as the use of *audible notifications*. This kind of feedback is not appropriate for every working environment, like open-space offices, but might be useful in noisy production areas (Korn, 2012). Gamified systems should offer functionalities enabling *social interaction and communication* between users, for instance, chats or forums, to foster social interaction and to increase participation in general (Chen and Pu, 2014; Walsh and Golbeck, 2014). Social interaction can be further intensified by, for instance, enabling gifting of *virtual goods* (Nakajima and Lehdonvirta, 2013). The CRM example also includes virtual goods, which can be purchased by spending coins earned within the application. Similarly, functionality to invite colleagues to join the system should be implemented to increase the adoption of gamified systems (Schacht and Schacht, 2012). When introducing a gamified system, Master-Ms should be implemented to ease *onboarding* of users, that is, the act of bringing new users into the system (Iosup and Epema, 2014). This can be accomplished by awarding users with achievements for initial usage (Barata *et al.*, 2013), providing instructions, rules, or hints (Hsu *et al.*, 2013), as well as by implementing tutorials (Iosup and Epema, 2014). *Reminders* of user's past behavior, for instance, a history of actions, can be helpful and ease future work (Munson and Consolvo, 2012; Snipes *et al.*, 2014). Gamified ES should support *easy, enjoyable, and exciting interaction concepts* by, for example, making use of an attractive user interface with stimulating visuals (Gnauk *et al.*, 2012; O'Donovan *et al.*, 2013). The design can *visually resemble existing games*, like Tetris (Korn, 2012), or incorporate *fantasy* elements to make the experience more emotionally appealing to users (Li *et al.*, 2012). However, the use of such concepts might be limited due to the complexity of ES and the need to ensure that work is approached seriously. Another Master-M identified in the context of system design is *meaning*, in a sense that users are convinced their work will produce something meaningful (Foong *et al.*, 2011; Gnauk *et al.*, 2012; Nicholson, 2012). Meaning for users can be created by directly showing them how their actions influence the success of the organization.

*Challenges*

Another aspect of creating gamified ES relates to implementing challenges. Challenging tasks in ES guide users by providing missions and rewarding them after successful completion. They give users a feeling of working towards a goal and support structuring tasks (Korn, 2012). To create challenging working environments, clear *goals* have to be defined and presented (Passos *et al.*, 2011). Considering the CRM example, one goal, for instance, refers to the creation of a new product. The formulation of clear goals leads to enhanced user performance, for instance, by being satisfied after fulfilling a goal (Bandura, 1993). In turn, increased satisfaction positively influences the performance on future tasks with similar goals. It is proposed that users should be allowed to set their own goals (Morrison and DiSalvo, 2014) or choose favored goals (Hsu *et al.*, 2013). In this context, *shadowing* is derived from racing games and describes a method where the users' goal is to compete against their own records (Korn *et al.*, 2012). Goals can be linked by telling users a background *story* (Halan *et al.*, 2010; O'Donovan *et al.*, 2013). Another way of creating challenges is using *time pressure* (Halan *et al.*, 2010; Cheong *et al.*, 2013). This, however, might not be appropriate when the focus is on ensuring qualitative work. Furthermore, the concept of 'Flow' is often described in relation with challenges. It refers to a state where the challenges that users face almost perfectly meet their skills (Nakamura and Csikszentmihalyi, 2001). In this state, users are neither bored nor overstrained. Combining the concept of flow and the identified Master-M *progressive disclosure* – which describes the adjustment of task difficulties based on the increasing skills of users (Fitz-Walter *et al.*, 2011; Li *et al.*, 2012) – gamified ES should provide challenges that automatically adjust their degree of difficulty, thus striving for a state of flow. Within the context of our CRM example, a first task might refer to accomplishing the initial login. Continuously, users are then confronted with further tasks like conducting contact searches and creating new products. Additionally, *access* – describing what users can see and do in the system – can be granted or increased after completing goals in order to foster progressive disclosure (Iosup and Epema, 2014). The use of challenges in ES might be limited due to a high number of monotone and standard tasks. Nonetheless, challenges can be used, for example, to motivate users to explore the ES in detail, to show unknown or new functions and features, as well as to demonstrate new ways of solving problems.

*Rewards*

Most Gamification applications, like the CRM example (see Figure 2), make use of *point systems* (Zichermann and Cunningham, 2011). Point systems reward users by adding a certain number of points to user accounts for completing actions or combinations of these (Burke and Hiltbrand, 2011). They motivate users due to their cumulative nature, which drives users to remain active (Burke and Hiltbrand, 2011; Smith, 2011). For a successful implementation of point systems in ES, transparency related to the purpose of a point system in general and the manner in which points are awarded must be provided (Nicholson, 2012). Transparent point systems form a foundation for many other Master-Ms, such as achievements, in addition to being easy to implement and integrate into existing systems. Previously, achievements and badges have been used as alternative forms of rewarding users. We define *achievements* as rewards for completing a clear and desirable goal (Liu *et al.*, 2011). In the CRM example, an exemplary achievement is the creation of an account. In contrast, *badges* consist of optional rewards and goals whose fulfillment is stored beyond the scope of core activities (Hamari, 2013). Some badges might be designed as hidden and are only awarded by surprise if some special actions are taken (Domínguez *et al.*, 2013). Additionally, users will be motivated to continue exploration of ES features in order to discover these hidden badges. Considering the CRM example, users are encouraged to explore the ES since they can expect to earn collection items (see the lower bar in Figure 2) in random hidden corners. On the one hand, engagement of users is additionally increased due to their desire to collect all available badges or achievements (Hsu *et al.*, 2013; Ibanez *et al.*, 2014). On the other hand, users' engagement in using the ES can be reduced after successfully collecting all rewards (Ibanez *et al.*, 2014), or users might get frustrated because of failing to gather all rewards (McDaniel *et al.*, 2012). To counteract this problem, gamified ES can implement rewards, which require repetition of tasks, or introduce new badges after a critical amount of users achieved all rewards. It is important to create meaningful, compelling, and challenging rewards to attract a broad range of users and foster long-term engagement (Nicholson, 2012; Haaranen *et al.*, 2014). Additionally, losing a badge induces an emotional impact, which is referred to as a user's tendency to *loss aversion* (Stockinger *et al.*, 2013). Achievements, badges, and point systems can be further combined with the Master-M *bonus*. This Master-M relies on rewards for having completed a series of challenges or tasks (Burke and Hiltbrand, 2011). In games, bonuses typically take the form of funny levels or additional game functions. To apply bonuses in gamified ES, they must fit into the underlying reward system. For instance, bonus points can be rewarded after successful completion of a special task or achievement (Fernandes *et al.*, 2012). Moreover, bonus mini games can be awarded in ES, after a completion of a series of tiring tasks, aiming to re-establish concentration and motivation.

*User Specifics*

The most often coded Master-M in this cluster is *user levels*. In video games, user levels document players' abilities and progress (Gnauk *et al.*, 2012). However, they can be easily transferred to more serious areas, such as Gamification in the ES context (see the lower left part of Figure 2). They can express expertise or skill maturity levels of an employee in specific fields. In the CRM example, the level represents the experience derived from the points earned within different areas (e.g., sales, account management) of the CRM system. Additionally, user levels can be used to support knowledge management, by enabling an electronic catalogue of employee skills (Lindvall and Rus, 2003; Garud and Kumaraswamy, 2005), thus offering a possibility to easily evaluate knowledge acquisition and dissemination. User levels can also be used to define goals and support progressive disclosure by changing the difficulty and offering new challenging goals after users reach a new level (Burke and Hiltbrand, 2011). Enabling users to move to a higher level can lead to a desire to earn reputation or status by reaching high levels, thus increasing motivation. Users should be able to maintain a *user profile* (e.g., the dashboard in Figure 2), which can be private (e.g., for showing current achievements and badges) or public (e.g., enabling comparison with colleagues and friends) (Ziesemer *et al.*, 2013; Morrison and DiSalvo, 2014). A *virtual character* can be understood as a virtual avatar representing an employee. This Master-M enables self-expression and can also be used to support the social dynamic of sharing virtual goods for characters or to represent a certain user level (Barata *et al.*, 2013). The upper left side in Figure 2 could include an avatar instead of showing a picture of the respective employee. Lastly, *ideological incentives* can as well be used to motivate employees (Nakajima and Lehdonvirta, 2013). It refers to influencing someone's attitudes and values with regard to a desired behavior, hence increasing intrinsic motivation.

## 4.2 Gamification Dynamics

In this section, we describe the Master-Ds that can result from using gamified ES equipped with the Master-Ms presented in the previous section. Besides describing the Master-Ds, we refer to selected Master-Ms in order to explain how they relate to increased user motivation and engagement.

By rewarding users (e.g., with badges), Master-D of *ownership* might develop. This concept represents a positive, sustained connection to an entity, for instance, a badge with visual representation leading to a feeling of shared ownership (Burke and Hiltbrand, 2011).

Badges for the successful creation of a new product within the CRM system are a good example. The badge can lead to the feeling of owning the product or rather its idea. Users might be motivated to sustain or expand their ownership, which leads to increased work performance. However, ownership can in some contexts lead to undesired consequences (e.g., grudge or discomfort) (Cramer *et al.*, 2011). Additionally, losing a badge induces an emotional impact, which is referred to as a user's tendency to loss aversion (Stockinger *et al.*, 2013). An example in ES is a badge that is awarded once a certain level of work quality is achieved, and is withdrawn if the level of quality decreases. In general, *competition*, for instance, resulting from Master-Ms like leaderboards (see Figure 2 for an example), can positively influence motivation of users. However, the introduction of competition should be carefully evaluated based on the context since a strong emphasis on competition can lead to decreased participation (Chen and Pu, 2014; de-Marcos *et al.*, 2014). Competition with other employees might not be appropriate in some working environments. As an alternative, shadowing can be implemented (see "Challenges"). In the CRM example, employees might compete against the points they achieved in the previous week or month. The opposite social Master-D of competition is *collaboration*. It rallies people to work together to solve problems and overcome challenges (Burke and Hiltbrand, 2011; Schacht *et al.*, 2014). Gamified ES that support collaboration might improve problem-solving processes, facilitate team synergy effects, and increase motivation. By strengthening relationships in collaborative teams, Master-D *altruism* can emerge. It refers to users' concern for the welfare of other members of their team. The aforementioned gifting of virtual goods (see "System Design") within the CRM example is one example for this Master-D. When evaluating collaboration in a gamified ES, a relevant Master-D is *social facilitation* (Hamari and Koivisto, 2013). It describes an effect where individual users achieve better results at simple tasks when working in a group or being in company of other individuals (Zajonc, 1965). Similarly, *conforming behavior* (i.e., the desire not to act against group consensus) (Nakajima and Lehdonvirta, 2013) and *group loyalty* can lead to increased participation (Hsu *et al.*, 2013). Most humans have a desire for *status* and *reputation*. These desires can be satisfied by humans performing specific actions themselves, thus increasing users' motivation to engage in these actions (Deterding, 2012; Vassileva, 2012; Vasilescu *et al.*, 2014). In particular, completion of achievements and personal level-ups (see "Rewards", "User Specifics" and the leaderboards, levels, and collections in Figure 2) as well as presenting received rewards to others will support satisfaction of these desires. In addition to the desire for status and reputation, employees might also *envy* other employees and desire to have something other employees have (e.g., achievements, badges, or bonuses) (Burke and Hiltbrand, 2011). However, this Master-D requires that an employee's achievements are visible to others. It should be noted that envy can foster competition, which is why its potential occurrence in workplace settings must be evaluated carefully as well. Another identified Master-D in this category is *self-expression*, which results from a desire to express autonomy, identity, or originality, or to mark one's personality as unique. While self-expression leads to increased participation and engagement, it might be difficult to realize this effect in the ES context. Possible use cases include offering the possibility to write work-related articles (Bista *et al.*, 2012) or to implement a billboard to foster communication.

### 4.3 Gamification Risks

Whereas we identified direct negative consequences of implementing Gamification in two studies only, we coded seven Master-Risks, which are shown in Table 4. We elaborate on these risks in the following.

**Table 4** Gamification Master-Risks.

| Master-Risk | Description | Reference |
|---|---|---|
| Suffering task quality | Quality of tasks might suffer if gamified elements distract from the main purpose of activities. | Blohm and Leimeister (2013) |
| Malfunction and failures | A low implementation quality of M&Ds might lead to malfunction and failures concerning reward systems or interaction concepts, which in turn reduce user motivation or lead to user frustration. | Yates and Wootton (2012), de-Marcos *et al.* (2014). |
| Cheating | If underlying rules are not clearly defined, it enables cheating, which can lead to rejection of implemented game elements by other employees. | Zichermann and Cunningham (2011), Reeves and Read (2009) |
| Privacy breach | Monitoring and surveillance of both the performed activity and the performing employee are likely to breach privacy rights. | Reeves and Read (2009) |
| Overemphasis of competition | An overemphasis of competition might lead to deceasing participation and not appeal to every employee. Competition might undermine cooperation, which is needed in business contexts. | de-Marcos *et al.* (2014) |
| Declining effects | A decreased effectiveness can occur once the novelty of Gamification has worn off. For instance, challenges might gradually be perceived as too simple. | Nakajima and Lehdonvirta (2013), Burke and Hiltbrand (2011) |
| Undermining intrinsic motivation | By excessively granting extrinsic rewards, the underlying intrinsic motivation can be undermined. | Haaranen *et al.* (2014), Hakulinen *et al.* (2013) |

One of the identified risks associated with Gamification relates to *task quality*, which can suffer when gamified elements distract users from the main purpose of activities (Blohm and Leimeister, 2013; Haaranen *et al.*, 2014). Productivity loss might be another consequence of such distraction. Consequently, it is imperative to implement an adequate level of Gamification. A low implementation quality of M&Ds can lead to *malfunction and failures* concerning reward systems or interaction concepts, which in turn reduce user motivation or lead to user frustration. If users build up aversion against low quality game elements, it is likely that they will not continue to use the ES. Similarly, many critics emphasize that current applications of Gamification often reduce the complexity of well-designed and balanced games to its simplest components, such as gathering points and rewarding badges (also referred to as 'Pointification') (Deterding, 2012; Vassileva, 2012). Hence, high quality of game elements and high ease of use have to be ensured when gamifying ES (Yates and Wootton, 2012; de-Marcos *et al.*, 2014). Moreover, it is likely that productivity declines when employees feel disadvantaged due to *cheating* of other employees (Reeves and Read, 2009; Zichermann and Cunningham, 2011; O'Donovan *et al.*, 2013). In the worst case, not clearly defined rules allow for cheating, which can lead to rejection by other employees. Gamification requires both clear rules and controls to prevent cheating (Reeves and Read, 2009; Thom *et al.*, 2012; Rapp *et al.*, 2012). In addition, *privacy issues* need to be considered. Gamification offers new ways of electronic monitoring and surveillance (Reeves and Read, 2009). In gamified applications, data can be collected for both, the activity performed and the employee performing the activity (e.g., ES usage, individual performance). Through such monitoring, employee privacy and personal rights are more likely to be harmed. Likewise, publication of the gathered data may lead to other negative consequences. For example, disseminating data concerning challenges (e.g., failures or slow progress) can decrease motivation or undermine the relationships between employer and employees (Reeves and Read, 2009). Such issues can be avoided through the differentiation between private and public data (Burke and Hiltbrand, 2011; Snipes *et al.*, 2014). Furthermore, employees should be given the choice whether to publish private data (Reeves and Read, 2009). Anonymized leaderboards are a means to preserve user privacy and prevent demoralizing users on the one hand, still enabling users to compare themselves to other employees in the leaderboard on the other hand (Halan *et al.*, 2010; Iosup and Epema, 2014). Considering the CRM example (see Figure 2), while an employee's rank within a leaderboard is shown, the employee does not get any information about the identity of the employees ranked below or above the own rank. Gamification also bears the risk that employees perceive a high level of organizational control (Nicholson, 2012; Yates and Wootton, 2012), which can result in feelings of autonomy loss and reduced self-control (Reeves and Read, 2009). If all information within the dashboard of the CRM example (see Figure 2) is visible to management, supervisors are enabled to control employees' work actions in detail. To counteract this hazard, individual data should only be used in an aggregated form (Nicholson, 2012; Yates and Wootton, 2012). Otherwise, Gamification's positive effects on employee motivation can be undermined by threats to trust in the employing organization. Furthermore, application of game elements, which foster competition, has to be carefully evaluated. On the one hand, competition can strongly foster participation and deepen engagement of employees. On the other hand, an *overemphasis of competition* can lead to decreasing participation (de-Marcos *et al.*, 2014) and is probably not appealing for all employees. Competition might as well undermine cooperation, which is needed in businesses contexts (Spencer, 2013). When promoting and rewarding collaboration, it is important to consider users who only participate occasionally, yet are essential for reaching collaborative goals (Spencer, 2013). Such users might not have a chance to climb leaderboards or receive badges, and "may feel that the top badge earners are devoting time to [Gamification] itself rather than to their jobs or even the actual goal of the project" (Spencer, 2013: 60). Hence, these users can easily get demotivated by a gamified system to participate in further collaborative tasks. Implementing Gamification should not only be about a meaningful design of game elements into work activities, but also incorporate the long-term perspective and organizational strategic objectives. Otherwise, an organization will not be prepared for the risk of *declining effects* over time, which might occur once the novelty of Gamification has worn off (Nakajima and Lehdonvirta, 2013; Chen and Pu, 2014). Employee skills will likely improve due to challenges accomplished (Passos *et al.*, 2011), which will consequently be perceived as too simple and require adaptations and changes of M&Ds to ensure continued Gamification benefits. This, however, requires additional expenditure over time. Another concern is the decrease of satisfaction due to continuous rewards of gamified elements in case of successful task accomplishments (Burke and Hiltbrand, 2011). Especially, Gamification has been criticized for undermining the *underlying intrinsic motivation* by granting extrinsic rewards (Haaranen *et al.*, 2014: 37). An incentive approach based on addressing users' personal needs might reduce such negative effects (Farzan and Brusilovsky, 2011). Moreover, the consequences of removing gamified elements need to be considered. For example, if implemented game elements are removed from an ES, employee performance can decline below the level prior to introducing Gamification (Nicholson, 2012; Thom *et al.*, 2012). The CRM example includes the earning of coins (see Figure 2) to acquire virtual goods. When removing such options, users are likely to discontinue usage since previous efforts are no longer acknowledged. Thus, complete removal of gamified elements should be carefully considered. Alternatively, a new gamified system can be implemented to keep motivation high (Thom *et al.*, 2012). This approach, however, requires continuous investments (Zichermann and Cunningham, 2011). In general, organizations face the risk of underestimating costs and expenditures when gamifying ES (O'Donovan *et al.*, 2013).

## 5. Discussion

Our study provides an overview of the various game elements pertaining to Gamification as well as the risks related to application of Gamification. Our review of extant literature on Gamification yielded four clusters of 27 Master-Ms as well as 11 Master-Ds that can be applied to gamify ES. We also revealed seven Master-Risks to be considered when gamifying ES. In the following, we discuss both practical and research implications of our study.

### 5.1 Practical Implications

Organizations should use Gamification to make ES usage more enjoyable for employees and as a means of fostering end-users' intrinsic motivation towards the task. By increasing the intrinsic motivation of end-users, organizations can improve acceptance and thus productivity of their ES. Our study makes a first step in that direction by providing a detailed description of Gamification Master-M&Ds that can be applied in the ES context. In the following, we provide implications on how organizations can use Gamification to address several barriers with regard to acceptance of ES by end-users. Organizations can address barriers of ES acceptance on the three levels: ES, organization, and end-users (see "Enterprise System Acceptance"). To counteract ES complexity and the low probability of end-users to actually use the ES (Kelley *et al.*, 2013; Bagayogo *et al.*, 2014), organizations should provide continuous training to enable end-users to accept the ES (Doll et al., 2003; Chang and Chou, 2011; Chou et al., 2014). For this purpose, Gamification can be applied by providing several training levels with increasing difficulty. For successfully accomplishing a level, *badges* can be awarded and new *levels* are unlocked (e.g., Paharia, 2013a, b; IBM, 2014). That way, organizations can counteract end-user abandoning an ES, in particular in the early post-adoptive phase (Schwarz *et al.*, 2014; Tong *et al.*, 2015).

Furthermore, companies should use the variety of game elements available to increase the intrinsic motivation of their ES end-users. Game elements such a *story*, *fantasy*, and *visually resembling existing games* can help end-users to identify themselves with the system. Ford, as an example, provides a gamified area called Ford p2p Cup (Paharia, 2013b), which is based on car racing. By using Gamification in this context, Ford attempts to increase the use of existing content, to accelerate personal certifications, and to motivate users to learn. *Badges*, for instance rewarded for watching videos or consuming latest product information, are issued in a virtual trophy cabinet.

By using game elements such as hidden badges, which are only awarded by surprise if some special actions are taken (Domínguez *et al.*, 2013), organizations can foster end-users to explore the ES. By striving to uncover the hidden badges, end-users are likely to explore an ES in order to increase their skills and knowledge towards the systems and to close the gap between actual system usage and the understanding of the system's full potential (Cooper and Zmud, 1990; Yi and Davis, 2003; Chang and Chou, 2011). Accordingly, end-users might discover innovative ways of ES usage to best support their tasks (Seddon *et al.*, 2010; Ke *et al.*, 2012).

A major contribution is the synthesis of Gamification M&Ds. We identified the main aspects that should be considered when applying or evaluating Gamification approaches. Since our focus is placed on the IS domain, we transfer Gamification to a novel context. Although we identified a positive influence of most game elements on employee motivation (see Online Appendix), Gamification should not be applied carelessly since negative influences were found in at least two contexts as well. First, in an educational context, students using a gamified e-learning system got lower participation scores (de-Marcos *et al.*, 2014). It was suggested that the gamified system emphasizes competition over collaboration and sharing, thus reducing student participation. Second, interviews revealed that the Master-D ownership could lead to undesired consequences (e.g., grudge or discomfort) (Cramer *et al.*, 2011). Besides carefully implementing Gamification, we believe that expedient designs need to consider the variety of Master-M&Ds identified in our review (see Appendices A and B). Our findings suggest that presence of interdependencies enables and amplifies the effectiveness of Gamification in the ES domain. An example of an interdependency is a leaderboard (Burke and Hiltbrand, 2011), which is usually implemented along with a point system (Zichermann and Cunningham, 2011).

While mechanics like leaderboards can easily be implemented, attention needs to be paid to their specific design. Due to misguided designs, leaderboards and respective competition can also prevent users from becoming engaged with the gamified system at all. If leaderboards are only designed for all-time scores, new users are unlikely to start competing since their chances to successfully compete for the top ranks are rather low. This example thus illustrates that both sides of the coin need to be considered. The sole implementation of a mechanic does not automatically lead to the desired dynamic. In order to avoid employees being discouraged by excessive competition, regular updates of leaderboards should be conducted (Depura and Garg, 2012; Gordillo *et al.*, 2013). Moreover, leaderboards for different business areas, groups of employees, or time intervals (see Figure 2) should be established to ensure that competition does not exclude specific users from being able to compete for the top ranks. To avoid failure of Gamification, organizations should carefully decide which game elements to integrate in their ES. For this purpose, several approaches have been proposed (Neeli, 2012; Werbach and Hunter, 2012; Chou, 2013). The three-step approach proposed by Neeli (2012), for instance, first includes assessing the main purpose of the task to be gamified. The second step involves identification of underlying objectives for different employees involved in task at hand (Aparicio *et al.*, 2012). Such objectives can, be derived from analyzing users' past behaviors (Blohm and Leimeister, 2013). Finally, game elements are selected to increase employee motivation towards reaching underlying objectives. Adequate combinations of M&Ds can yield synergetic effects. Thus, a systematic and expedient approach might lead to a higher success rate of Gamification in ES contexts.

## 5.2  Research Implications

Gamification as an approach to foster motivation of ES end-users is related to research on technology acceptance (Youngberg *et al.*, 2009; Codish and Ravid, 2014). As such, Gamification can be used as a means to increase the motivation to use technology. Thus, first attempts have been made to analyze Gamification's effect on technology acceptance, showing that "gamification objectively yields improvements in factors, such as software enjoyment, flow experience or perceived ease of use" (Herzig *et al.*, 2012: 803). The general impact notwithstanding, analysis of the effect of specific M&Ds on technology acceptance constructs is subject to future research. Investigations might refer to the analysis of whether reception of badges increases users' self-efficacy concerning the ES (Davis *et al.*, 1989). Considering the CRM example, badges for rather simple tasks like system logins and contact searches (see Figure 2) might increase users' perceived ease of use, increase users' self-efficacy concerning the usage of the CRM system, and thus contribute to continued system usage.

Considering the broad stream of research concerning technology acceptance in general and ES acceptance in particular, previous studies have predominantly analyzed which factors contribute to end-users acceptance of ES (Nwankpa and Roumani, 2014a). The reasons for how and why end-users accept ES helps organizations to understand (lack of) acceptance. Yet, the reasons do not provide means of how to overcome the barriers. By providing game mechanics and respective dynamics, Gamification helps organizations to guide their ES end-users behavior. As an instance of the gamified CRM example in this study, we perceive *competition* (see "Gamification Dynamics") as a typical Master-D that this triggered by Gamification's Master-M *leaderboard* (see "Gamification Mechanics"). The leaderboard serves the purpose of showing users their rank among other users in the gamified CRM systems. Implementing a leaderboard requires to specify an underlying measurement system to assess and compare users' positions. Typically, the amount of awarded points, badges, achievements, and fulfilled goals as well as the user level are used as a measurement system. The sole existence of a leaderboard within the gamified system can make employees engage in system usage. Motivation to compete with others is caused by the joy of being the 'leader' or at least being in front of direct colleagues. People's compulsion to climb the leaderboard results in continuous competition and system usage.

While we provide both mechanics and dynamics, the relation between Master-Ms and Master-Ds has not been analyzed yet. Within our literature analysis, we did not identify major trends between Master-Ms and specific Master-Ds. While the use of point systems and leaderboards likely leads to competition rather than collaboration, in-depth insights into such dependencies are yet to be discovered and might depend on the types of players among users (see our discussion of different player types below). Investigating long-term effects of Gamification is promising as well, especially considering the reduced positive effects of Gamification over time (see "Gamification Risks"). Accordingly, it appears worthwhile to analyze whether badges granted for a specific number of logins (see Figure 2) have a motivating effect on employees in the long run. Empirical investigations should hence focus on corroborating the identified influences as well as uncovering interdependent effects of Gamification Master-M&Ds. While Gamification has already been effectively applied as a motivator in areas like health (Hamari and Koivisto, 2013), research (Parra et al., 2013) and education (Akpolat and Slany, 2014; Li *et al.*, 2013), attention should be paid to the setting in which Gamification is applied (Blohm and Leimeister, 2013). Although we identified Master-M&Ds for Gamification of ES in general, Gamification is not suitable for each context and not every Master-M&D can be effectively applied to all settings (Hamari, 2013). It is thus essential to fully comprehend the context before developing a Gamification design (Rapp *et al.*, 2012).

Likewise, two further aspects to be considered are employees' affinity for games and the novelty of the ES in which Gamification is applied (Hamari, 2013). In view of the former, personality is considered a factor to impact the way Gamification is perceived (Codish and Ravid, 2014). Employees' affinity for (video) games is crucial for Gamification effectiveness and will vary among employee groups, especially concerning different generations. We believe that a workspace with mostly young employees is the most promising environment for Gamification, as older employees' affinity to digital games might not be sufficient (McGonigal, 2011). While young employees are likely to be familiar with features like the hidden items in the CRM example (see "Rewards"), such features might exclude older employee generations from such areas of the gamified system. Regarding the characterization of ES users, for whom Gamification is implemented, previous research has identified four player types (Bartle, 1996; Schacht and Schacht, 2012). First, *explorers* enjoy understanding and exploring the game world. Within the CRM example, such players might, for instance, be suitably addressed by hidden items that are likely to be collected by exploring different areas of the CRM application. Second, *achievers* are eager to complete the majority of the challenges with which they are confronted. Such players might be eager to reach high user levels in the CRM example. Third, *socializers* get involved in games mainly due to other players like themselves. Badges that are awarded for jointly accomplished tasks (e.g., new products created by a team of CRM employees) might appeal to this player type. Finally, *winners* strive for the accomplishment of challenges at the expense of other players (i.e., they perceive a challenge as good if there can be one winner only). This player type is likely to be addressed by badges that are only award once (e.g., the first user to create ten new products). Since each player type reacts on and is motivated by implemented M&Ds in different ways, we suggest assessing which type of player motivation is predominant among ES users in a given context. Future research should thus address the dependencies between types of player motivation and effective implementation of Gamification.

## 6. Conclusion

We present Gamification as an innovative approach to foster user motivation to utilize ES. Considering Gamification as a means to overcome the various barriers to ES acceptance, we guide organizations in their attempt to increase the acceptance of technology, in particular when considering future generations of ES users, who are considered to have a high affinity to video games. In addition to the general motivation to use Gamification in organization contexts, we illustrate the various options to design Gamification applications. Our synthesized Master-M&Ds are of particular interest since they show the comprehensive potential to motivate employees in general, and ES users in particular, by implementing Gamification. Providing a comprehensive list of Master-M&Ds, we aim to counteract the reduction of meaningful Gamification to simple 'Pointification'. Designers might use this synthesis as a starting point for developing, selecting, combining, and evaluating applicable M&Ds for future systems. Finally, our presentation and discussion of risks related to the application of Gamification provide insights into its limitations. Organizations should not see Gamification as a 'magic bullet' for increasing user motivation, but as a means that has to be carefully and deliberately integrated into organizational structures to support a motivational culture.

# 7. Appendix A – Gamification's Master-Ms

**Table 5** Identified Master-Ms.

| Name (# coded) | Description |
|---|---|
| *System Design* | |
| Audible Notifications (4) | Implementing sound effects and/or background music (Li *et al.*, 2012). |
| Exciting interaction concepts (7) | "This includes an attractive user interface with stimulating visuals and exciting interaction concepts, as well as a high degree of usability" (Gnauk *et al.*, 2012: 105). |
| Fantasy (2) | "Fantasy evokes images of objects or situations that aren't actually present. This can make the experience more emotionally appealing to users" (Li *et al.*, 2012: 105). |
| Feedback (25) | Immediate feedback is used to keep players aware of their progress or failures in real-time (Passos *et al.*, 2011). |
| Leaderboards (29) | "[...] Leaderboards are used to track and display desired actions, using competition to drive valuable behavior" (Bunchball Inc., 2010: 10). |
| Meaning (9) | "[…] For meaningful gamification, it is important to take into consideration the background that the user brings to the activity and the organizational context into which the specific activity is placed. […] The game elements need to come out of aspects of the underlying activity that are meaningful to the user" (Nicholson, 2012: 2-5). |
| Onboarding (3) | The act of bringing new users into the system, for example, by providing tutorials (Iosup and Epema, 2014). |
| Redo (2) | A user is allowed to attempt activities a number of times if unsuccessful, or to achieve sub-goals (Tootell *et al.*, 2014). |
| Reminder (4) | Reminder of past behavior of the user, for instance, a history of actions (Liu *et al.*, 2011). |
| Social interaction functions (11) | Offering functionalities to enable interaction and communication between users, for instance, chats or forums. |
| Virtual goods (5) | Virtual goods are non-physical, intangible objects that can be purchased or traded (Bunchball Inc., 2010). |
| Visually resembling existing games (3) | Creating a visual design that is similar to existing games. For example, designing the system similar to the well-known Tetris game (Korn, 2012: 315). |
| *Challenges* | |
| Access (3) | Access describes what users can see and do inside the system, and might be granted or increased after completing goals (Iosup and Epema, 2014), especially to foster progressive disclosure. |
| Goals (28) | Goals of the underlying activity should be adapted as challenges for the user (Passos *et al.*, 2011). |
| Progressive disclosure (7) | "A game helps players to continuously increase their skills by progressive disclosure of both knowledge and challenge […]. This will help ensure that the challenges in the game match the player's skill levels […]" (Li *et al.*, 2012: 105). |
| Shadowing (3) | Shadowing describes a method where users attempt to improve their previous records (Korn *et al.*, 2012). |
| Story line (6) | "A story line links the tasks together to create a cohesive whole" (Villagrasa and Duran, 2013: 430). |
| Time pressure (8) | Creating time pressure on activities, for instance, through counters or hourglasses (Li *et al.*, 2012). |
| *Rewards* | |
| Achievement (23) | A reward for completing a clear and desirable goal (Liu *et al.*, 2011). |
| Badges (19) | "Badges consist of optional rewards and goals whose fulfilment is stored outside the scope of the core activities of a service" (Hamari, 2013: 2). |
| Bonus (8) | Bonuses are rewarded for having completed a series of challenges or core functions (Burke and Hiltbrand, 2011). |
| Loss aversion (3) | Loss aversion is a game mechanic that influences user behavior not by a reward, but by instituting punishment when the targeted goal is not achieved (Liu *et al.*, 2011). |
| Point system (29) | Point systems reward users for completing actions, whereby a numeric value is added to their overall points total (Burke and Hiltbrand, 2011). |
| *User Specifics* | |
| Ideological incentives (1) | "[...] Ideological incentives is the notion of influencing user behavior through influencing their attitudes and values, in other words, educating the user on a deeper level. The ideological incentive makes it possible to motivate the user by himself" (Nakajima and Lehdonvirta, 2013: 11). |
| User levels (14) | "Levels indicate the proficiency of the player in the overall gaming experience over time [...]" (Gnauk *et al.*, 2012: 104-105). |
| User profile (5) | A profile, showing information about the user (e.g., awarded badges). |
| Virtual character (6) | An avatar representing the employee (Passos *et al.*, 2011). |

# 8. Appendix B – Gamification's Master-Ds

**Table 6** Identified Master-Ds.

| Name (# coded) | Description |
| --- | --- |
| Altruism (5) | In this context, altruism refers to virtual gift giving with the aim of strengthening the relationships between users (Nakajima and Lehdonvirta, 2013). |
| Collaboration (12) | "The community collaboration game dynamic rallies an entire community to work together to solve a riddle, resolve a problem, or overcome a challenge" (Burke and Hiltbrand, 2011: 13). |
| Competition (17) | Contests enable users to challenge each other (Bunchball Inc., 2010). |
| Conforming behavior (5) | "Conforming behavior is the desire not to act against group consensus, colloquially known as peer pressure" (Nakajima and Lehdonvirta, 2013: 117). |
| Envy (1) | This dynamic is based on the user's desire to have what others have (Burke and Hiltbrand, 2011). |
| Group loyalty (1) | Group loyalty represents users' affective and cognitive allegiance to the group as users participate in a group (Ibanez *et al.*, 2014). |
| Ownership (5) | "The ownership dynamic represents a positive, sustained connection to an entity that leads to a feeling of shared ownership" (Burke and Hiltbrand, 2011: 14). |
| Reputation (8) | "Reputation is based on the opinion of other users about the user or her contribution" (Vassileva, 2012: 183). |
| Self-expression (5) | Self-expression results from having a desire to express autonomy, identity or originality, or to mark one's personality as unique (Bunchball Inc., 2010). |
| Social facilitation (3) | Describes an effect where individual users achieve better results at simple tasks in the presence of other people or when working in groups (Zajonc, 1965). |
| Status (10) | "[…] Status can be earned by the user in isolation, by performing certain actions" (Vassileva, 2012: 183). |